\documentclass[twoside,12pt]{article}
\usepackage{epsfig}

\newcommand{\be}{\begin{equation}}
\newcommand{\ee}{\end{equation}}
\newcommand{\bea}{\begin{eqnarray}}
\newcommand{\eea}{\end{eqnarray}}

\begin{document}

\title{\vspace{1cm}Single- and Double-Pion Production in Nucleon Collisions
  on the Nucleon and on Nuclei -- 
the ABC Effect and its Possible Origin in a Dibaryonic Resonance}
\author{H.\ Clement, M.\ Bashkanov, E.\ Doroshkevich,
O.\ Khakimova, \\ F.\ Kren, A. \ Pricking, T.\ Skorodko and G.J.\ Wagner \\
\\ for the CELSIUS-WASA Collaboration \\
\\
Physikalisches Institut der Universit\"at T\"ubingen, \\
    Auf der Morgenstelle 14, D-72076 T\"ubingen, Germany\\
}
\maketitle
\begin{abstract}
The ABC effect -- an intriguing low-mass enhancement in the $\pi\pi$ invariant
mass spectrum -- is known from inclusive measurements of two-pion production
in nuclear fusion reactions. First exclusive measurements carried out at
CELSIUS-WASA for the fusion reactions leading to d or $^3$He reveal this
effect to be a $\sigma$ channel 
phenomenon associated with the formation of a $\Delta\Delta$ 
system in the intermediate state and combined with a resonance-like behavior
in the total cross section. Together with the observation that the
differential distributions do not change in shape over the resonance region
the features fulfill the criteria of an isoscalar s-channel resonance in $pn$
and $NN\pi\pi$ systems, if the two emitted nucleons are bound. It obviously is
robust enough to  survive in nuclei as a dibaryonic resonance
configuration. In this context also the phenomenon of $N\Delta$ resonances is
reexamined. 
\end{abstract}
\section{Introduction}

The ABC effect -- first observed by Abashian, Booth and Crowe \cite{abc} in
the double pionic fusion of deuterons and protons to $^3$He -- stands for an
unexpected enhancement at low masses in the spectrum of the invariant
$\pi\pi$-mass $M_{\pi\pi}$. Follow-up experiments
\cite{hom,hal,bar,ban,ban1,plo,abd,col,wur,cod} revealed this
effect to be of isoscalar nature and to show up in cases, when the two-pion
production process leads to a bound nuclear system. With
the exception of low-statistics bubble-chamber measurements \cite{bar,abd} all 
experiments conducted on this issue have been inclusive measurements carried
out preferentially with single-arm magnetic spectrographs for the detection
of the fused nuclei.

Initially the low-mass enhancement had been interpreted as due to an unusually
large $\pi\pi$ scattering length and as evidence for the $\sigma$ meson,
respectively \cite{abc}. Since the effect showed up particularily clearly at
beam energies corresponding to the excitation of two $\Delta$s in the nuclear
system, the ABC effect was interpreted lateron by a $\Delta\Delta$
excitation in the course of the reaction process leading to both a low-mass 
and a high-mass enhancement in isoscalar $M_{\pi\pi}$ spectra
\cite{ris,barn,anj,gar,mos}. In fact, the
missing momentum spectra from inclusive measurements have been in support of
such predictions. It has been shown \cite{alv} that these structures can be
enhanced considerably in theoretical calculations by including $\rho$ exchange
and short-range correlations.


\section{Experiment}

In order to shed more light on this issue, exclusive
measurements of the reactions $pd \rightarrow pd\pi^0\pi^0$ ($T_p$ = 1.03 and
1.35 GeV) and  $pd \rightarrow ^3$He$\pi\pi$ ($T_p$ = 0.893 GeV) 
have been carried out in the energy region of the ABC effect at CELSIUS using
the 4$\pi$ WASA detector setup with pellet target system \cite{zab}.
The selected energies have been close to the maximum of the ABC effect
observed in the respective inclusive measurements. The $pd \rightarrow
pd\pi^0\pi^0$ reaction proceeds as quasifree $pn \rightarrow d\pi^0\pi^0$
reaction with a spectator proton of very small momentum in the lab
system. Since all ejectiles with the exception of the spectator have been
measured, the spectator momentum has been reconstructed by kinematical fits
with three overconstraints. Preliminary results for the reaction can be found
in recent conference proceedings \cite{inpc,menu1,menu2,erice1}.
The experimental results on the $pd \rightarrow
^3$He$\pi^0\pi^0$ and $pd \rightarrow ^3$He$\pi^+\pi^-$ reactions have been
published already in Ref. \cite{bash,MB}.

\section{Experimental Results}

Some specific results of the CELSIUS-WASA measurements are shown in Figs. 1
and 2 for
the double-pionic fusion to the deuteron, which is the most basic 
reaction for studying the ABC-effect.  Fig. 1 depicts 
the spectra of the invariant masses $M_{\pi^0\pi^0}$ and  $M_{d\pi^0}$ for the
quasifree $pn \rightarrow d\pi^0\pi^0$ reaction at the beam energy $T_p$ = 1.03
GeV
\footnote{
Note that due to Fermi motion of
the nucleons in the target deuteron  the quasifree reaction process proceeds
via a continuum of effective collision energies in the range 0.94 - 1.18
GeV with according kinematical smearing in the differential
distributions. This smearing may be reduced strongly by dividing the data into
narrow bins of effective collision energy at the cost of statistics.} 
. 
 
The $\pi^0\pi^0$ channel, which is free of any isospin I=1 contributions,
exhibits a pronounced low-mass enhancement (ABC effect) in the
$M_{\pi^0\pi^0}$ spectrum both in the  
fusion process to the deuteron and in the one leading to
$^3$He\cite{bash,MB}. We note that in the $^3$He$\pi^+\pi^-$ channel the 
threshold enhancement is observed \cite{bash} too, however, less
pronounced. The reason for this is that this channel in addition contains 
isovector contributions - as may be seen  \cite{colin} by the small shifts
between the $\Delta$ peaks in the $M_{^3He\pi^+}$ and $M_{^3He\pi^-}$
spectra -- see Fig. 5 of Ref. 
\cite{bash}. However, the main result of these measurements is that indeed
two $\Delta$s  are excited simultaneously in this reaction -- in support of the
hypothesis that a $\Delta\Delta$ system is excited in the course of the
double pionic fusion process.


\begin{figure}
\begin{center}
\begin{minipage}[t]{16.5 cm}
\includegraphics[width=0.45\textwidth]
  {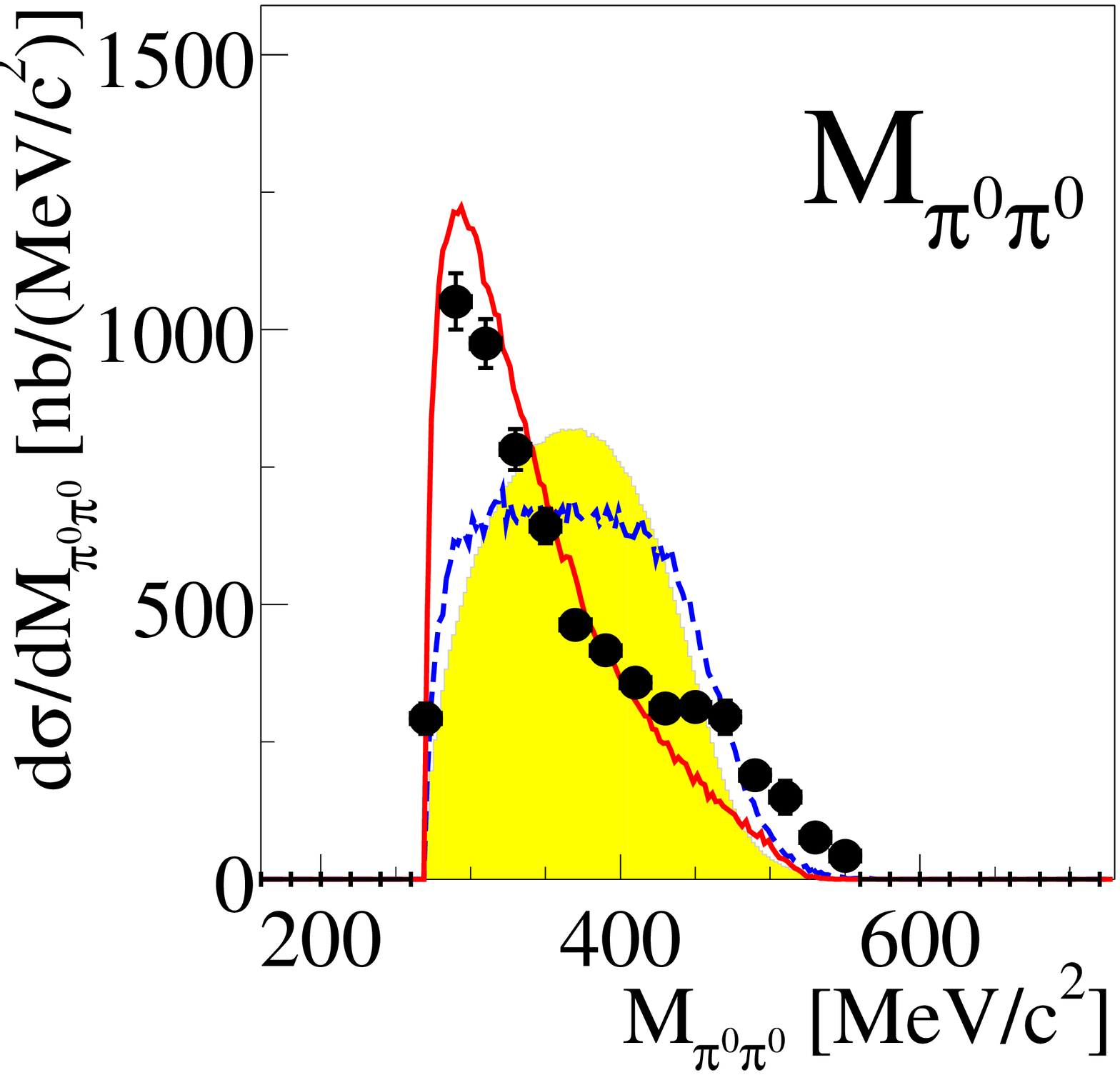}
\includegraphics[width=0.45\textwidth]
    {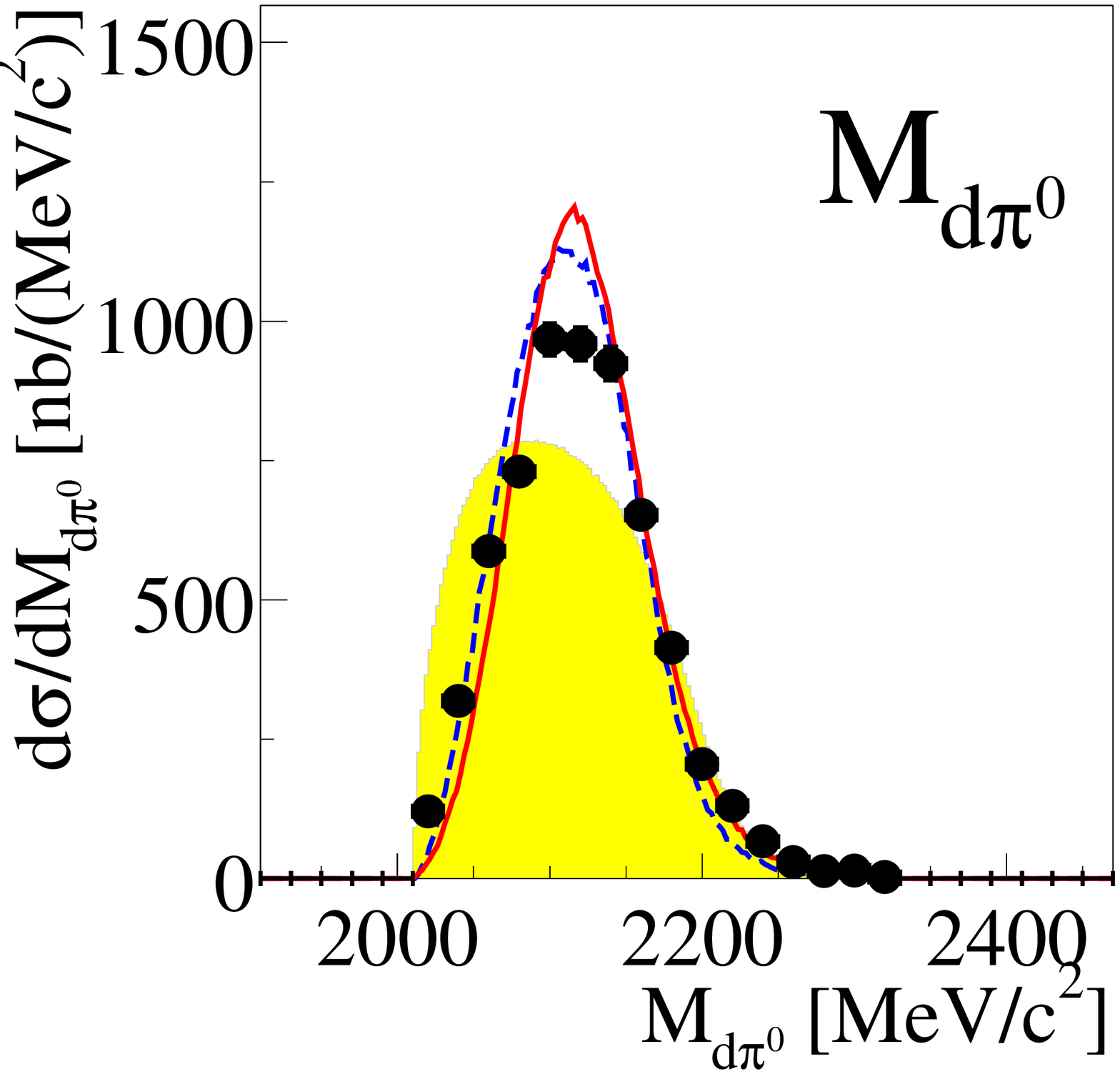}

\end{minipage}
\begin{minipage} {16.5 cm}
\caption{
  Distributions of the invariant masses $M_{\pi^0\pi^0}$ and 
  $M_{d\pi^0}$ from the exclusive measurement
  of the quasifree $pn \to d \pi^0\pi^0$ reaction at a beam energy $T_p$ =
  1.03 GeV. The shaded areas show the pure phase 
  space distributions. Solid and dashed curves give
  $\Delta\Delta$ calculations with and without the assumption of a  quasibound
  state in the 
  $\Delta\Delta$ system leading to a resonance in the $pn$ and $d\pi^0\pi^0$
  systems (from Ref. \cite{menu1}).
  }
\end{minipage}
\label{fig2}
\end{center}
\end{figure}


From the measured angular distributions \cite{menu1,bash,MB} we find the
following features:
\begin{itemize}
\item The pion angular distribution in the $\pi\pi$ subsystem is flat for the 
  low-mass enhancement region in the $M_{\pi\pi}$ spectrum, i.e., the
  ABC-effect is of scalar-isoscalar nature -- in other words it is a $\sigma$
  channel phenomenon.
\item The distribution of the opening angle between the two pions shows that
  the ABC-effect is associated with two pions leaving the interaction vertex
  in parallel.
\item The angular distribution of the $\pi\pi$ system (which is aquivalent to
  the angular distribution of the residual nucleus) in the overall
  center-of-mass system is not flat. It rather corresponds to a double p-wave
  distribution as expected from the decay of the $\Delta\Delta$ system.
\end{itemize}. 

From the measurements of the quasifree $pn \to d\pi^0\pi^0$ reaction at
various energies we notice that the differential distributions do not change
in shape significantly with energy. This points to the dominance of a single
partial wave in the entrance channel, as is the case for the excitation of a
s-channel resonance. As a consequence of such an assumption we should find a
resonance-like energy dependence in the total cross section. Actually this is
what in fact is observed for this reaction.
In Fig. 2 we show the energy dependence of the total cross section of
the double-pionic fusion to the deuteron. Depicted are the results for the $pn
\to d\pi^+\pi^-$ reaction from bubble chamber measurements at DESY \cite{bar}
and JINR \cite{abd} together with the preliminary CELSIUS-WASA results
\cite{inpc,menu1} for 
the quasifree $pn \to d\pi^0\pi^0$ reaction at two incident energies, which
have been binned into narrow ranges of effective collision energy providing
thus four entries below and two entries above the peak energy. Since
$\pi^+\pi^-$ and $\pi^0\pi^0$ channels are related by an isospin factor of
two, the $\pi^0\pi^0$ results are plotted in Fig. 2 multiplied by this isospin
factor. A resonance-like behavior of the total cross section is obvious.

\section{Discussion and Interpretation of Experimental Results}

The $\pi\pi$ low-mass enhancements observed in the exclusive data for the  
$\pi^0\pi^0$ channels turn out to be much larger than predicted in previous
$\Delta\Delta$ calculations\cite{ris,anj,mos}. As an example we show by the 
dashed lines in Fig.1 and by dotted lines in Fig. 2 calculations in the
model ansatz of 
Ref. \cite{ris}, where we additionally included the pion angular 
distribution in $\Delta$ decay and  the Fermi smearing of the nucleons bound
in the final nucleus. Contrary to these predictions the data also do not
exhibit any high-mass enhancement in $M_{\pi^0\pi^0}$ that had been supported by
the inclusive measurements, too. As suspected already in Ref. \cite{col} the 
high-mass bump observed in inclusive spectra rather turns out to be associated
with $\pi\pi\pi$ and $\eta$ production as well as with I=1 contributions. 
Since on the one hand the available $\Delta\Delta$ calculations 
obviously fail, but on the other hand the data clearly show the $\Delta\Delta$
excitation in the $M_{N\pi}$ spectra, a profound physics piece appears to be 
missing. Such a missing piece may be provided by a strong $\Delta\Delta$
attraction or even a boundstate formation, as was demonstrated in
Refs.\cite{bash,MB,mb,OK}. 

\begin{figure} [t]
\begin{center}
\begin{minipage}[t] {16.5 cm}
\includegraphics[width=1.0\textwidth]{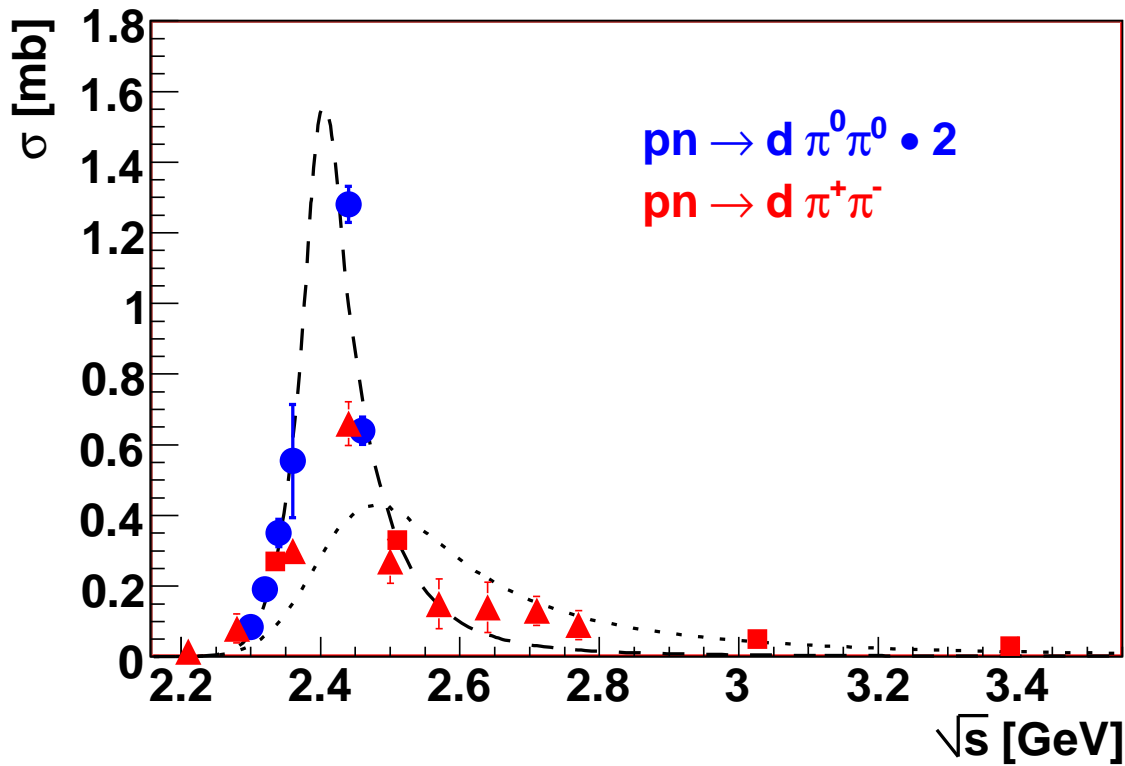} 
\end{minipage}
\begin{minipage}[t] {16.5 cm}
 \caption{Energy
   dependence of the $pn \to d\pi\pi$ reaction with preliminary results of
   this work for the $\pi^0\pi^0$ channel ( quasi-free measurements at two
   incident energies) and results for the $\pi^+\pi^-$
   channel from Ref. \cite{abd}(squares) and Fig. 2c of Ref. \cite{bar} 
   (triangles). Dashed and dotted lines represent calculations with and
   without the 
   assumption of a quasibound state in the 
  $\Delta\Delta$ system leading to a resonance in the $pn$ and $d\pi^0\pi^0$
  systems.
}
\end{minipage}
\end{center}
 \end{figure}

The essential clue to the nature of the ABC effect appears to be in
the intriguing energy dependence of the double-pionic fusion in the isoscalar
channel. We note that the isovector fusion channel $pp \to d\pi^+\pi^0$,
shows {\bf no} ABC effect \cite{FK} despite a clear 
$\Delta\Delta$ excitation signal in its differential spectra. It also exhibits
an energy dependence \cite{bys} in its total cross section, which is close to
the dotted curve in Fig. 2. In contrast the isoscalar
fusion channel exhibits a much more pronounced energy-dependence in accordance
with a resonance excitation having a width of roughly
100 MeV or even less, i.e. much smaller than twice the $\Delta$ width
expected from usual $\Delta\Delta$ calculations. As also borne out by the
data in Fig. 2 the cross section maximum at
$\sqrt{s} \approx$ 2.4 GeV means that the resonance mass is below twice the
$\Delta$ mass, i.e. a quasibound state in the isoscalar $\Delta\Delta$
system, which not only can decay into the $pn$ system, but also into the
isoscalar $d\pi\pi$ system, because the $\Delta$ decay width is larger than
the binding of this state.

In fact, if we use a 
Breit-Wigner term with a $q_{\Delta\Delta}$ dependent width and adjust the
width parameters not to fit the total cross section data, but to 
reproduce the ABC-effect in the $M_{\pi^0\pi^0}$ spectra, then we obtain not
only a quantitative description of all
differential data (see, e. g., solid curves in Fig. 1) but at the same time also a
quantitative description of the energy dependence of the total cross section
(dashed curve in Fig. 2)thus obtaining automatically the observed width of the
total cross section data.  

Before proceeding here any further, we would like to discuss the
experimentally necessary conditions to establish a s-channel resonance. In
case of a two-body reaction these may be given easily as follows:
\begin{itemize}
\item a resonance must be governed by a single partial wave and
\item this partial wave must exhibit a resonant behavior in its real and
  imaginary amplitude parts leading to a pronounced looping in the Argand plot
  of this particular partial wave.
\end{itemize}

That way and by detailed partial wave analyses also resonances can be sensed, 
which in the observables are not seen directly, because they are
burried underneath a background of other processes. Even in the case that a
single partial wave contains also a lot of nonresonant contributions, a pole
search in the complex plane of the partial-wave amplitude as well as an Argand
plot can reveal hidden resonances -- as is well known, e. g., from
the study of the Roper resonance \cite{said,TS}.

If in a particular reaction excitation and decay of a resonance is the
dominating process over a
sufficiently large energy region, then the situation
is much simpler:
\begin{itemize}
\item the total cross section must exhibit a Breit-Wigner like energy
  dependence and
\item the angular distibution has to keep the same shape over the region of the
  resonance ("frozen distribution") being symmetric about 90$^{\circ}$ in the
  center-of-mass system.
\end{itemize}

\subsection{Resonances in the $N\Delta$ system}

In order to examine this situation  on an example close to our problem, we
reinvestigate the situation of $N\Delta$ resonances, which show up both
in elastic pp and $\pi d$ scattering, and, in particular, in the $pp \to
d\pi^+$ reaction, which also has been measured over a wide energy range
\cite{saida}. 
For all three reaction channels there are ample data in the region of
interest. Also there exist detailed partial wave analyses for each of
these reaction channels separately \cite{saida,saidb,saidc} as well as
combined analyses \cite{saidd}. Whereas elastic NN scattering
couples to many channels and hence is quite insensitive to particular
resonance states, elastic $\pi d$ scattering and especially the reaction
$pp \to d\pi^+$ select specifically single $\Delta$ excitations in two-body
processes. The latter reaction has been investigated intensively at the pion
factories LAMPF, TRIUMF and PSI via the reversed reaction $\pi^+ d \to pp$. 

The data for the $pp \to d\pi^+$ reaction exhibit
an energy dependence of the total cross section, which is in very close
correspondence to a Breit-Wigner resonance peaking at $T_p \approx$ 550
MeV and sitting on a very low background at higher energies. In addition 
the angular distibution, which has a $(3~cos^2\Theta_{cm} + 1)$ dependence -- 
specific to $\Delta$ excitation and decay --, keeps its shape from a few
MeV above threshold up to $T_p \approx$ 650 MeV. I. e., this reaction shows
already on a qualitative level all features of a s-channel resonance, which
carries the characteristics of a single $\Delta$ excitation.

With such features in the data it is not surprising that their partial-wave
analysis exhibits a clear resonant behavior in the partial waves $^1D_2P$,
$^3F_3D$, $^3P_2D$ and $^1G_4F$ (in the notation
$^{2S_{pp}+1}L^{pp}_JL^{\pi}$). All four partial waves  
describe textbook examples of a resonant behavior in real and imaginary parts
of their amplitudes with close-to-perfect loopings in the Argand plot (see
Figs. 7 and 8 of
Ref. \cite{saida}). The by far dominating partial wave is $^1D_2P$, which
governs more than 90$\%$ of the total cross section up to $T_p \approx$ 500
MeV (see Fig. 5 of Ref. \cite{saida}). At higher energies also the other three
partial waves come into play. Their partial wave cross sections peak right at
the $N\Delta$ 
mass of $\sqrt{s}$ = 2170 MeV. The cross section of the  $^1D_2P$
partial wave, however, peaks already at lower energy corresponding to
$\sqrt{s}$ = 2130 MeV. The resonant cross sections have a width of roughly 120
MeV, which conforms with the width of the $\Delta$ resonance. The special role
of the $^1D_2$ partial wave has been already noted as early as 1968 by R. N. 
Arndt in his article about the "Unbound Diproton" \cite{arndt}

All four partial waves represent resonant $N\Delta$ configurations. However,
only $^1D_2P$ corresponds to a configuration, where the relative orbital
angular momentum between $N$ and $\Delta$ is zero. This explains its
dominant role in this reaction.
Also it is the only configuration, which resonates roughly 40 MeV below the
$N\Delta$ threshold. This means that the isovector s-wave
interaction between $N$ and $\Delta$ is 
strongly attractive with the ability to form a I(J$^P$) = 1(2$^+$) state
with a binding of about 40 MeV. Since the width of the $\Delta$ constituent in
this state is larger than its binding energy, this state appears to be only
quasi-bound. We note, however, that despite this binding the total width of
this resonant state is obviously not significantly smaller than that of the
free $\Delta$ resonance. 
This result will be of importance for the discussion of resonant states in the
$\Delta\Delta$ system in the following paragraph. 

We finally note that without the assumption of a s-channel resonance the
experimental result for the angular distribution to coincide with that
for $\Delta$ excitation and decay would not be easily understandable. In case
of t-channel $\Delta$ excitation on one of the nucleons by meson exchange, the
relevant reference axis for the polar angle of the pion emitted in 
$\Delta$ decay would be given by the momentum transfer $\vec{q}$ and not by
the beam axis. Since the direction of $\vec{q}$ varies with the scattering
angle, the resulting cms-angular distribution for the pions with the beam
axis as the reference axis, would be smeared out yielding approximately a much
flatter $(1~cos^2\Theta_{cm} + 1)$ dependence.

\subsection{Quasibound State in the $\Delta\Delta$ System}

After the excursion to the $N\Delta$ system we come back to the discussion of
the $\Delta\Delta$ system. We see that similar to the
resonant $^3F_3D$, $^3P_2D$ and $^1G_4F$ partial waves, which peak right at
the $N\Delta$ threshold, the $\Delta\Delta$ excitation in the isovector channel
$pp \to d\pi^+\pi^0$ peaks right at the $\Delta\Delta$ threshold as expected
from a calculation of the $\Delta\Delta$ process without mutual interaction
of the two $\Delta$s. Also the experimentally observed width of the
resonance-like energy dependence of the total cross section coincides with the
expected width of twice the $\Delta$ width.

The situation is different for the isoscalar channels $pn \to d\pi^+\pi^-$
and  $pn \to d\pi^0\pi^0$, where the data suggest a peak postion near
$\sqrt{s}$ = 2.41 GeV, i. e. some 50 MeV below the $\Delta\Delta$
threshold. Here the situation is reminiscent of the $^1D_2P$ partial wave
in case of the
$N\Delta$ system. Also the resonance-like behavior in the total cross section
and the observed "frozen distributions" in the differential cross sections
fulfill the conditions of a s-channel resonance in these isoscalar channels.
Note that with three particles in the exit channel 
there are now much more differential distributions than just the single
angular distribution in case of only two ejectiles.

However, the big difference now is that we again observe a
width in the resonant cross section of 100 MeV, which is, however, only half
of that expected naively. In case of the  $^1D_2P$ partial wave in
the 
$N\Delta$ system the binding did not lead to a noticeable change of the decay
width. Hence we conclude that the observed width in the isoscalar channel is
obviously not just the simple result of the binding between the two $\Delta$
states. It 
rather signals more complicated configurations in the wave function of the
intermediate state, which hinder its decay -- as would be expected for a
non-trivial dibaryon state.

Our model calculations for the quasibound state in the $\Delta\Delta$
system assume that the decay into the  $\Delta\Delta$ system proceeds via
relative s-waves between the two $\Delta$s. Since these calculations describe
the measured 
angular distributions very well, we conclude that the spin-parity assignment
to this isoscalar intermediate state should be either $J^P$ = $1^+$ or $3^+$,
where we take into account that the two-fermion system has to be in an
antisymmetric state.

We note that the existence of such states has been predicted in different
theoretical calculations \cite{ping,oka,kuk} and sometimes \cite{ping}
referred to as "inevitable dibaryon".

\section{Conclusions and Outlook}

The finding of a s-channel resonance in the isoscalar $pn \to d\pi\pi$ channel
has the consequence that this resonance should show up also in the elastic
$pn$ scattering channel. Unfortunately in the corresponding energy region of 
$T_p$ = 1.0 - 1.3 GeV such data are very sparse or non-existent,
respectively. Moreover, from the analysis of the $pn \to d\pi\pi$  data
we expect the s-channel 
resonance to show up in $pn$ elastic scattering with a peak cross section 
of several 100 $\mu$barn only, which has to be compared with a total elastic
$pn$ cross section of about 20 mb in this energy region. I.e., only a detailed
partial wave analysis of very precise elastic $pn$ scattering data
 over the energy region $T_p$ = 1.0 - 1.3 GeV would have the potential to
 sense this resonance in the $^3S_1$ and $^3D_3$ partial waves, respectively.

Another finding from the analysis of the $pn \to d\pi^0\pi^0$ data is that
the ABC effect -- a $\sigma$ channel low-mass enhancement in the $M_{\pi\pi}$
spectra --  is intimately connected to the appearence of the s-channel
resonance in isoscalar $pn$ and $\Delta\Delta$ systems. Since the ABC-effect
shows up also in heavier nuclear systems, this means that this
resonance is a quite stable object, which even survives in the nuclear
surroundings. In fact, the energy dependence measured for the $^3$He and
$^4$He cases in previous inclusive measurements \cite{ban1} is in support of
this conclusion. 

The observed strong energy dependence of the $pn \to d\pi^0\pi^0$ reaction
certainly needs further experimental verification. This is even more true for
the energy dependencies observed for the double-pionic fusion reactions
leading to $^3$He and $^4$He. Therefore we have proposed to investigate
these energy dependencies in dedicated exclusive measurements at COSY using
there the newly installed and upgraded WASA detector, which recently had been
moved succesfully from the CELSIUS-ring at Uppsala to the COSY-ring at
Forschungszentrum J\"ulich.

\section*{Acknowledgments}

We acknowledge valuable discussions with R. A. Arndt, I. Strakovski, E. Oset,
C. Wilkin, Ch. Hanhart, 
A. Sibirtsev, L. Alvarez-Ruso, A. Kaidalov, V. I. Kukulin, L. Dakhno and
V. Anisovich on 
this issue. This work has been supported by BMBF (06TU201, 06TU261), COSY-FFE, 
DFG (Europ. Graduiertenkolleg 683), Landesforschungsschwerpunkt
Baden-W\"urttemberg and the Swedish Research Council. We also acknowledge the
support from the European Community-Research Infrastructure Activity under FP6
"Structuring the European Research Area" programme (Hadron Physics, contract
 number RII3-CT-2004-506078).

\end{document}